\documentclass[12pt]{article}
\usepackage{graphicx}
\usepackage{cite}
\usepackage{color}
\def\unit{\mbox{\large \boldmath $1$}}

\newcommand{\bm}[1]{\mbox{\boldmath $#1$}}
\newcommand{\fnd}[2]{\frac{\textstyle #1}{\textstyle #2}}
\newcommand{\x}[1]{{\textstyle #1}}
\newcommand{\abs}[1]{\left| #1\right|}
\newcommand{\fndrs}[4]{\fnd{\raisebox{#1}{$#2$}}{\raisebox{#3}{$#4$}}}
\newcommand{\xrm}[1]{{\textstyle \mbox{\rm #1}}}
\newcommand{\dissum}[2]{\displaystyle \sum_{#1}^{#2}}
\newcommand{\Real}[1]{\Re {\it e}\left(#1 \right)}
\newcommand{\Imag}[1]{\Im {\it m}(#1 )}

\newcommand{\bracket}[2]{\mbox{$\left\langle #1\left| #2\right.\right\rangle$}}
\newcommand{\braket}[3]{\mbox{$\left\langle #1\left|
#2\right| #3\right\rangle$}}
\begin{document}
\title{$S$-wave and $P$-wave $\pi\pi$ and $K\pi$ contributions\\ to
three-body decay processes\\ in the Resonance-Spectrum Expansion}
\author{
Eef van Beveren\\
{\normalsize\it Centro de F\'{\i}sica Te\'{o}rica,
Departamento de F\'{\i}sica, Universidade de Coimbra}\\
{\normalsize\it P-3004-516 Coimbra, Portugal}\\
{\small http://cft.fis.uc.pt/eef}\\ [.3cm]
\and
George Rupp\\
{\normalsize\it Centro de F\'{\i}sica das Interac\c{c}\~{o}es Fundamentais,
Instituto Superior T\'{e}cnico}\\
{\normalsize\it Universidade T\'{e}cnica de Lisboa, Edif\'{\i}cio Ci\^{e}ncia,
P-1049-001 Lisboa, Portugal}\\
{\small george@ajax.ist.utl.pt}\\ [.3cm]
{\small PACS number(s): 13.25.Ft, 13.75.Lb, 12.40.Yx, 12.39.Pn}
}

\maketitle

\begin{abstract}
We study $S$- and $P$-wave $\pi\pi$ and $K\pi$ distributions for the three-body
production processes $J/\psi\to\omega\pi^{+}\pi^{-}$,
$D^{+}\to K^{-}\pi^{+}\pi^{+}$,
$D^{+}\to K^{+}\pi^{+}\pi^{-}$, and $D^{+}_{s}\to K^{+}\pi^{+}\pi^{-}$,
by applying the Resonance-Spectrum Expansion to the two-body subamplitudes.
The results are compared to the corresponding signals reported by
the BES, E791 and FOCUS collaborations.
We find that the studied $S$- and $P$-wave channels can largely account
for the observed structures in the data up to 1 GeV.
\end{abstract}

\section{Introduction}

Experimental data on hadronic resonances have largely shifted from elastic
scattering to production processes over the past decade.
Nevertheless, the thus observed hadrons are supposed to be identical
to the corresponding ones first seen in elastic scattering.
This generally accepted uniqueness of resonance mass and width,
irrespective of the underlying process, reflects itself
in the universality of the pole structure of the resonance
amplitude or subamplitude.
However, this does of course not mean that a resonance is produced
with a fixed probability in all processes where it shows up.
This will depend on microscopic details of a specific production process,
the corresponding vertex functions, and possibly also on effects from
final-state interactions with other particles produced alongside.

Clearly, theory has not managed to keep up
with this fast experimental development.
This is no surprise in view of the extremely difficult dynamics
at the quark level involved in a three-meson decay, not to speak
of decays to even more mesons.
Nevertheless, it would already be significant progress if
one could use an amplitude ansatz
with the correct resonance-pole structure,
as derived from a demonstrably working model for elastic meson-meson
scattering and meson spectroscopy.

In this spirit, we propose to start from the
so-called Resonance-Spectrum Expansion (RSE)
in order to obtain such an ansatz.
For that purpose, we construct here the amplitude for the production of
a pair of interacting mesons out of a quark-antiquark pair,
using a previously developed coupled-channel model
\cite{PRD21p772}, and its reformulation \cite{IJTPGTNO11p179}
in terms of an expansion in the radial confinement spectrum
of the $q\bar{q}$ pair, the RSE.
We assume thereby that the decay processes
$J/\psi\to\omega\pi^{+}\pi^{-}$,
$D^{+}\to K^{+}\pi^{+}\pi^{-}$ and $D^{+}_{s}\to K^{+}\pi^{+}\pi^{-}$
take place in cascades of two subsequent $q\bar{q}$ pair-creation processes,
triggered by a weak process
and resulting in one spectator meson plus a pair of rescattered mesons.
For the decay
$D^{+}\to K^{-}\pi^{+}\pi^{+}$,
we assume that one of the $\pi^{+}$ mesons is formed in the
weak process, whereas the remaining $K^{-}\pi^{+}$
pair stems from $q\bar{q}$ pair creation.
In either case,
the spectator meson is supposed to be formed in the first step
of the cascade process, alongside a $q\bar{q}$ pair.
The latter pair is supposed to couple to the pair of
rescattered mesons.

Accordingly, we substitute the three-meson vertex for the final
process by the transition potential of Ref.~\cite{PRD21p772}.
This interaction couples $q\bar{q}$ to a pair of mesons.
This pair is supposed to interact via quark loops,
which is described by the RSE ladder sum.
The quark loops, representing the dynamics of the two interacting mesons,
can be written in a closed analytic form
\cite{PRD21p772,IJTPGTNO11p179},
and, consequently, the full production amplitude as well.
This is not the case for the initial three-meson vertex ---
either weak or strong ---
which is here just represented by a complex coupling constant.

The RSE has the advantage that all possible resonances
for a given set of quantum numbers are contained in one expression.
Therefore, the overlapping of resonances and their possible mixing
\cite{PRD64p094002}
is automatically taken care of.
Also, by employing spherical Bessel and Hankel functions
for meson-meson distributions, the expressions become automatically
real below threshold, thus curing
the problem of subthreshold complex phases occurring in other methods
\cite{ZPA359p173}.
But it has yet another advantage.
The elastic two-meson ($MM$) scattering amplitude,
which can be written in the form \cite{IJTPGTNO11p179,AIPCP687p86}
\[T={\cal V}+{\cal V}G_{f}{\cal V}+\cdots \; , \]
where $\cal V$ represents the Born term of the transition potential
$V(q\bar{q}\to MM)$ and $G_{f}$ the two-meson propagator,
leads for production to an expression of the form
\[\unit+G_{f}{\cal V}+G_{f}{\cal V}G_{f}{\cal V}+\cdots\;=\;{\cal V}^{-1}T\;.\]
As a result, we obtain for production, in an analytically exact manner,
just the common denominator of the elastic scattering amplitude
for all coupled scattering channels.
Hence, the pole structure of elastic scattering
is, with our method, manifestly preserved in production.
Of course, resonance line shapes can be different for elastic
scattering and production, since those also depend on the numerators
of the corresponding RSE expressions.

The three-meson vertex for the final process, here represented by the
transition potential of Ref.~\cite{PRD21p772}, results, through Bauer's
formula \cite{Bauer}, in a form factor determined by a spherical Bessel
function, which depends on the angular momentum of a specific two-meson
channel. This automatically yields the form factors and centrifugal
barriers that other approaches have to provide empirically (see e.g.\
Refs.~\cite{BlattWeisskopf,NPB471p59,HEPEX0411042}).

The initial vertices for strong cascades might be treated as well within the
recoupling scheme of Ref.~\cite{ZPC21p291}, but we shall leave this for later
study.  Here, the initial three-meson vertices are just represented by
complex coupling constants, as in the isobar model (see e.g.\
Ref.~\cite{NPB471p59} for a practical application), to be
adjusted to the data. Nevertheless, it should be mentioned that from the
recoupling scheme of Ref.~\cite{ZPC21p291} one obtains values for the
initial-process coupling constants which are very similar in magnitude to
the ones used here. This indicates that the cascade assumption is very
reasonable, but most probably that also some dynamics has to be considered
for the initial processes, before all can be predicted by theory.
In the prsent paper, it is our aim to demonstrate that the final production
vertex can be handled in a very elegant way through the RSE.

We shall compare our results to data taken by the
BES \cite{PLB598p149},
E791 \cite{PRL89p121801} and
FOCUS \cite{PLB601p10}
collaborations.
It turns out that the data can be very well described by our amplitudes.
However, contrary to the FOCUS analysis
\cite{PLB601p10}, we shall find here that the background in their data
is negligible.
In Refs.~\cite{AIPCP619p63,PLB632p471} there are similar conclusions
with respect to the data of E791 \cite{PRL89p121801},
LASS \cite{NPB296p493} and BESII \cite{PLB633p681}.
It is very fortunate that such data exist nowadays.
Twenty years ago it would have been impossible to verify
whether results as presented here make any sense at all.
Today, we may even discover that data are very well described.

The organisation of this paper is as follows.
In section~\ref{RSE1}, we discuss the details of the RSE
for elastic scattering.
The production amplitude for the RSE is derived
in section~\ref{Swavestudy},
and the $S$-wave predictions are compared to the data.
In order to improve the description of the data via RSE amplitudes,
we show that at least also $P$-waves must be included, which is
discussed in section~\ref{Pwavestudy}.
In section~\ref{RSE2} we give some more discussion of the details
of the RSE amplitudes for elastic scattering.
Section~\ref{conclusie} contains our conclusions.

\section{The Resonance-Spectrum Expansion}
\label{RSE1}

The RSE has been developed \cite{IJTPGTNO11p179}
for the description of meson-meson scattering resonances and bound states,
in a non-perturbative approach that aims at unquenching the $q\bar{q}$
confinement spectrum.
It consists of a simple analytic expression which can be
straightforwardly applied to all possible non-exotic systems of two mesons.
The RSE goes beyond simple spectroscopy, since it describes the scattering
amplitude, not only at a resonance, but also for energies where no resonance
exists. In contrast to models which have to rely upon numerical methods of
solution, the RSE has the additional advantage that the pole structure of its
scattering amplitude can be studied in great detail, owing to its closed
analytic form. The expression for the amplitude can even be analytically
continued, in an exact manner, to below the lowest threshold, where bound
states show up as poles on the real energy axis. The RSE easily handles many
coupled meson-meson channels, or coupled systems with different internal
flavours. As such, it is an ideal expression for the study of scattering theory
in general, i.e., the study of resonance structures and their relation to
some of the many $S$-matrix singularities, as well as the concept of Riemann
sheets and analytic continuation anywhere in the complex energy plane.

The basic ingredients of the RSE are confinement and quark-pair creation.
In its lowest-order approximation, a permanently confined quark-antiquark
system is assumed, having a spectrum with an infinite number of bound states,
related to the details of the confining force. We shall denote the energy
levels of this confinement spectrum by
$E_{n\ell_{q\bar{q}}}$ ($n,\ell_{q\bar{q}} =0$, 1, 2, $\dots$).
Here , we assume that the confinement spectrum is given by
\begin{equation}
E_{n\ell_{q\bar{q}}}\; =\; E_{0\ell_{q\bar{q}}}\, +\, 2n\omega
\;\;\; ,
\label{t00HO}
\end{equation}
which choice is anyhow rather immaterial for the purpose of our present study.
The parameter $E_{0\ell_{q\bar{q}}}$ represents, to lowest order,
the ground-state mass of the $\ell_{q\bar{q}}$-wave
of the quark-antiquark system,
which is related to the effective flavour masses
of the system. The strength of the confinement force is parametrised
by $\omega$ and gauged by the level splittings of the system.
In experiment one cannot directly measure the quantities
$E_{0\ell_{q\bar{q}}}$ and $\omega$,
because of the large higher-order contributions.
Consequently, the lowest-order system is purely hypothetical.
Nevertheless, one can obtain some order-of-magnitude insight
by examining mesonic spectra with more than one experimentally known
recurrency.

From the $J^{PC}=1^{--}$ charmonium and bottomonium states, one may conclude
that the average level splitting is of the order of 380 MeV, leading to
$\omega =0.19$ MeV, independent of flavour.
The latter property is compatible with the flavour blindness of QCD,
confirmed by experiment \cite{PRD59p012002}.
Indeed, the level splittings of the positive-parity $f_{2}$ mesons
seem to confirm that flavour independence can be extended
to light quarks \cite{HEPPH0610199}.
From the ground states of the recurrencies one may then extract
the order of magnitude of the effective quark masses,
e.g.\ $2m_{c}=m(J/\psi )=3.1$ GeV (in the RSE \cite{PRD27p1527}
we find for twice the effective charm mass the value 3.124 GeV),
or $2m_{u/d}=m(\rho )=0.77$ GeV (0.812 GeV in the RSE).
For the choice (\ref{t00HO}) of confinement force,
we determine
$E_{0\ell_{q\bar{q}}} = m_{q} + m_{\bar{q}} + (1.5 + \ell_{q\bar{q}} ) \omega$.
Once the effective flavour masses and $\omega$ are fixed \cite{PRD27p1527},
we may describe other systems, like scalar mesons and mixed flavours
\cite{ZPC30p615,PRL91p012003,HEPPH0312078,MPLA19p1949,PRL97p202001}.

Through quark-pair creation the $q\bar{q}$ system is coupled
to those two-meson systems which are allowed by quantum numbers.
In principle, many different two-meson channels can couple to
one specific quark-antiquark system. Here, since we study the properties
of the channel lowest in mass, we will strip the RSE of all other possible
two-meson channels, thereby assuming that their influence far below their
respective thresholds will be negligible.
Via consecutive quark-pair creation and annihilation,
a $q\bar{q}$ pair may also couple to pairs of different flavour,
for instance $u\bar{u}(d\bar{d})\leftrightarrow s\bar{s}$.
Here, as we study pion-pion scattering near threshold,
we shall assume that the coupling of a light pair of flavours
to strange-antistrange can be ignored.

The intensity of quark-pair creation is in the RSE parametrised by
the flavour-independent parameter $\lambda$. In principle, it has to be
adjusted to the data.  However, one may get an idea of the right order of
magnitude by the following reasoning. For small values of $\lambda$, one may
determine the width of the ground-state resonance in the one-channel case
(pion-pion here) by \cite{AP105p318}
\begin{equation}
\Gamma\;\approx\;
\lambda^{2}\, E_{0}\,
\fndrs{8pt}
{\sin^{2}\left(\fndrs{2pt}{a}{-3pt}{2}\sqrt{E_{0}^{2}-4m^{2}_{\pi}}\,\right)}
{-5pt}
{\fndrs{2pt}{a}{-3pt}{2}\sqrt{E_{0}^{2}-4m^{2}_{\pi}}}
\;\;\; ,
\label{RSEwidth}
\end{equation}
where $a$ represents the average distance at which light quark pairs are
created, and which can also be defined in a flavour-independent fashion
\cite{PRD21p772}. Note that this picture of a relatively well-defined
interquark distance at which pair creation is favoured qualitatively agrees
with very recent unquenched lattice simulations of string breaking
\cite{PRD71p114513}. Even our values of $a$, at least for heavy quarks,
are roughly in agreement with the latter lattice predictions.
For light quarks, $a$ is about 0.6 fm, as we will see
later on. If we take, for example, the $f_{0}$(1370) resonance width of
0.2--0.5 GeV, then we obtain $\lambda\approx$ 0.60--0.95.
This is of the order of 1, which would not allow the approximation
(\ref{RSEwidth}). Nevertheless, we actually employ here a value for $\lambda$
which is of the same order of magnitude as our estimate
(see caption of Fig.~\ref{t00Figures}).

\section{The RSE production amplitude
for the \bm{I=0} \bm{\pi\pi} \bm{S}-wave}
\label{Swavestudy}

The RSE amplitude suitable for our purposes results from the ladder sum in
quark-pair creation \cite{IJTPGTNO11p179},
and has for $\pi\pi$ elastic scattering the form
(${\cal V}$ and $G_{f}$ represent respectively the
quark-loop box and the two-pion propagator.)
\begin{eqnarray}
\lefteqn{T\left({\vec{p}\,},{{\vec{p}\,}'\,}\right)\, =\,
\braket{{\vec{p}\,}}{\left\{
{\cal V}+{\cal V}G_{f}{\cal V}+\dots\right\}}{{\vec{p}\,}'\,}}
\nonumber \\ [.3cm] & = &
\fnd{1}{8\pi^{2}}\,
\sum_{\ell =0}^{\infty}\, (2\ell +1)
P_{\ell}\left(\hat{p}\cdot{\hat{p}\,}'\;\right)\,
\fndrs{15pt}
{2\lambda^{2}\mu pa\, j_{\ell}(pa)\, j_{\ell}(p'a)\,
\dissum{n=0}{\infty}
\fndrs{3pt}{g_{n\ell}^{2}}
{-3pt}{E_{n1}-E(p)}}
{-15pt}{1-2i\lambda^{2}\mu pa\,
j_{\ell}(pa)\, h^{(1)}_{\ell}(pa)\,
\dissum{n=0}{\infty}
\fndrs{3pt}{g_{n\ell}^{2}}
{-3pt}{E_{n1}-E(p)}}
\;\;\; ,
\label{Tfull}
\end{eqnarray}
where $E_{n\ell}$ is defined in Eq.~(\ref{t00HO}), and where
\begin{equation}
E(p)\; =\; 2\sqrt{p^{2}+m_{\pi}^{2}}
\;\;\;\;\xrm{and}\;\;\;\;
\mu\; =\;\mu(E)\; =\;\frac{1}{4}E
\;\;\; .
\label{kandmuinE}
\end{equation}
The factors $g_{n\ell}$ are remnants \cite{ZPC21p291}
of the quark-antiquark distributions associated
with the confinement spectrum (\ref{t00HO}).
The amplitude (\ref{Tfull}) satisfies the unitarity condition
$\abs{1+2iT}=1$ for all energies $E>2m_{\pi}$,
as can be easily verified.
\begin{figure}[htbp]
\begin{center}
\begin{tabular}{c}
\resizebox{0.47\textwidth}{!}{\includegraphics{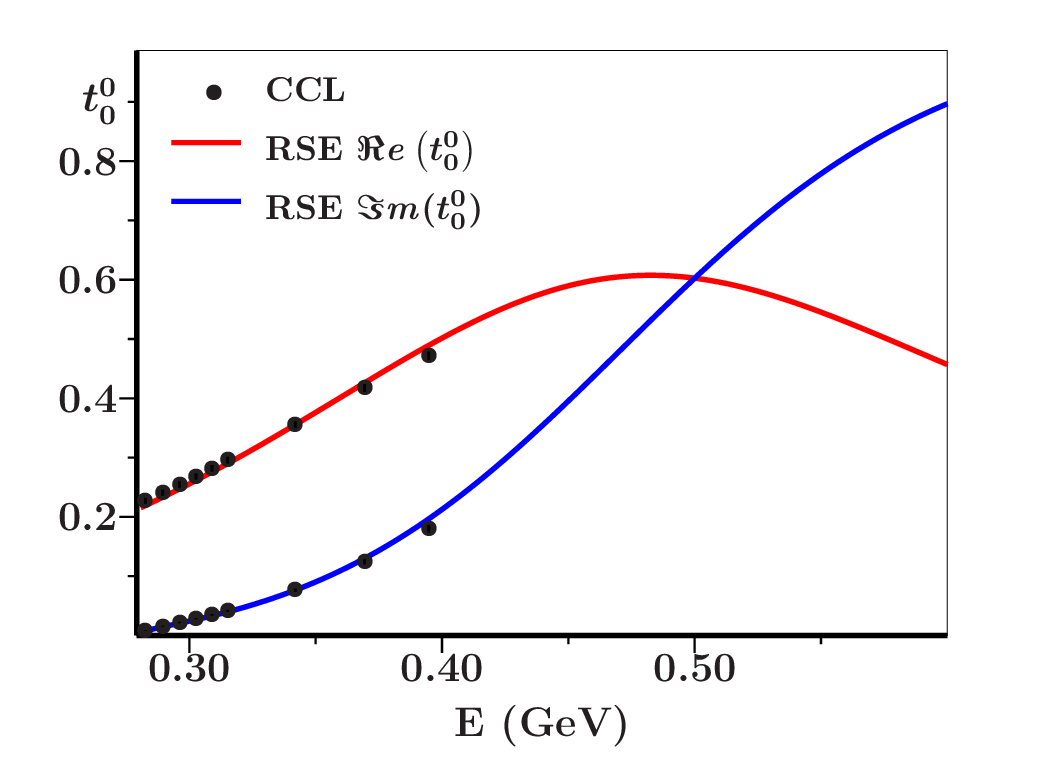}}\\ [-20pt]
\end{tabular}
\end{center}
\caption[]{\small $\Real{t_{0}^{0}}$ (red curve) and $\Imag{t_{0}^{0}}$
(blue curve) for the $S$-wave term of the RSE of Eq.~(\ref{Tfull}),
compared with the results of Caprini, Colangelo and Leutwyler
(CCL) \cite{PRL96p132001} ($\bullet$).
The RSE parameters are $\lambda =  1.29$, $a =  2.90$ GeV$^{-1}$,
$E_{01} =  1.30$ GeV and $\omega =  0.19$ GeV.
The recoupling parameters are here given by
$g^{2}_{n0}=(n+1)\times 4^{-n}$.}
\label{t00Figures}
\end{figure}

In Fig.~\ref{t00Figures} we compare the RSE amplitude
with the dispersion-relation result of Caprini, Colangelo and Leutwyler
\cite{PRL96p132001}.
We do not distinguish here between neutral and charged pions,
and take the pion mass equal to the one of the charged pions.
At threshold we find for the RSE
\begin{equation}
t_{0}^{0}\left( E=2m_{\pi}\right)\;
=\;\Real{t_{0}^{0}}\left( E=2m_{\pi}\right)\;
=\; 0.212
\;\;\; ,
\label{t00atthreshold}
\end{equation}
which may be compared to data at $0.220\pm 0.005$
\cite{NPB603p125}.

In Fig.~\ref{t00Figures}
one observes from the behaviour of both
$\Real{t_{0}^{0}}$ and $\Imag{t_{0}^{0}}$ that the RSE clearly describes
a resonant structure, i.e., the $f_{0}$(600) alias $\sigma$ meson.
However, for energies above 600 MeV,
the RSE prediction does not follow the data.
The main reason is the absence of a coupling to $s\bar{s}$ in the $q\bar{q}$
sector, as well as the $K\!\bar{K}$ channel (see e.g.\ Ref.~\cite{PLB641p265}).
However, for energies below 400 MeV
the agreement with the data is excellent.
For more details, see Ref.~\cite{HEPPH0702117}.

For production processes, we need to consider the amplitude
\cite{PPNP45p157,PRC73p045208,PLB639p59}
\begin{eqnarray}
\lefteqn{A\left(\vec{p}\,\right)\, =\,
\bracket{n\bar{n}}{\pi\pi}\, =}
\nonumber\\ [10pt] & & =
\int d^{3}k\,\braket{n\bar{n}}{V}{\vec{k}\,}
\left\{
\bracket{\vec{k}\,}{\vec{p}\,}\, +\,
\braket{\vec{k}\,}{G_{f}{\cal V}}{\vec{p}\,}\, +\,
\braket{\vec{k}\,}{G_{f}{\cal V}G_{f}{\cal V}}{\vec{p}\,}\, +\,\dots
\right\}
\nonumber\\ [10pt] & & =
\int d^{3}k\,\braket{n\bar{n}}{V}{\vec{k}\,}\,
\braket{\vec{k}\,}{1+G_{f}T}{\vec{p}\,}
\, =\,
\int d^{3}k\,\braket{n\bar{n}}{V}{\vec{k}\,}\,
\braket{\vec{k}\,}{{\cal V}^{-1}T}{\vec{p}\,}
\;\;\; ,
\label{ToVfull}
\end{eqnarray}
where $V$ represents the transition potential for $q\bar{q}\to\pi\pi$,
which, in the model of Eq.~(\ref{Tfull}), is represented by
a local delta-shell potential \cite{IJTPGTNO11p179},
and where ${\cal V}=V^{T}\left[ H_{n\bar{n}}-E\right]^{-1}V$
represents the quark-loop box.

\begin{figure}[htbp]
\begin{center}
\begin{tabular}{cc}
\resizebox{0.47\textwidth}{!}{\includegraphics{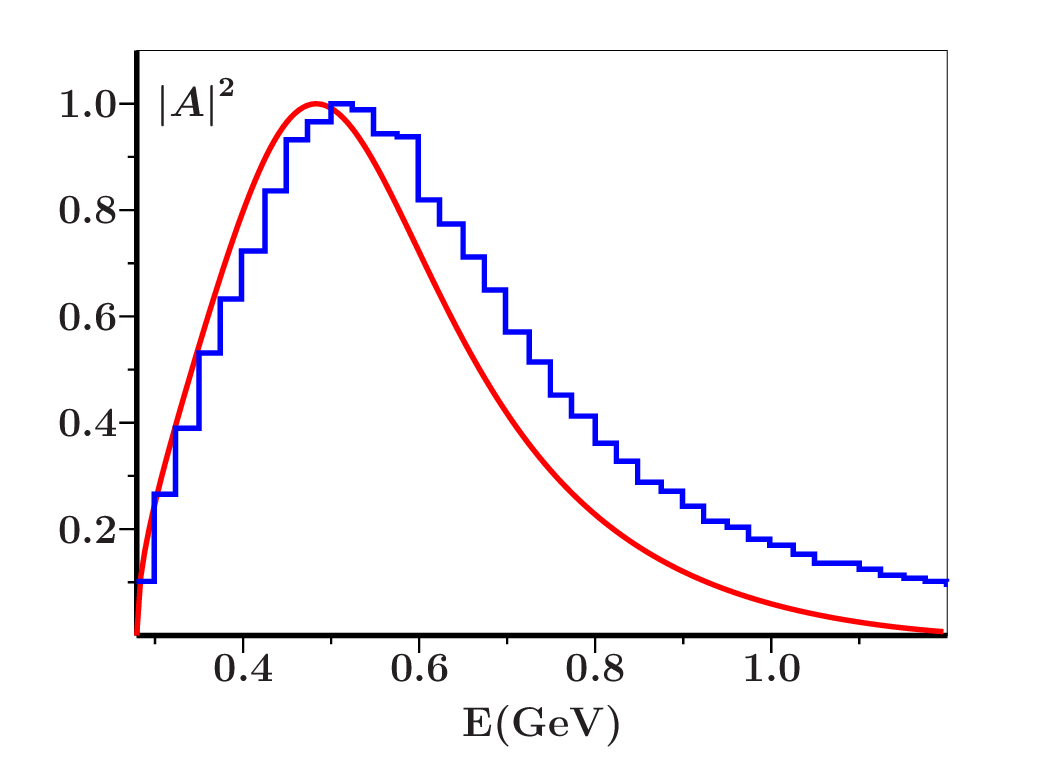}} &
\resizebox{0.47\textwidth}{!}{\includegraphics{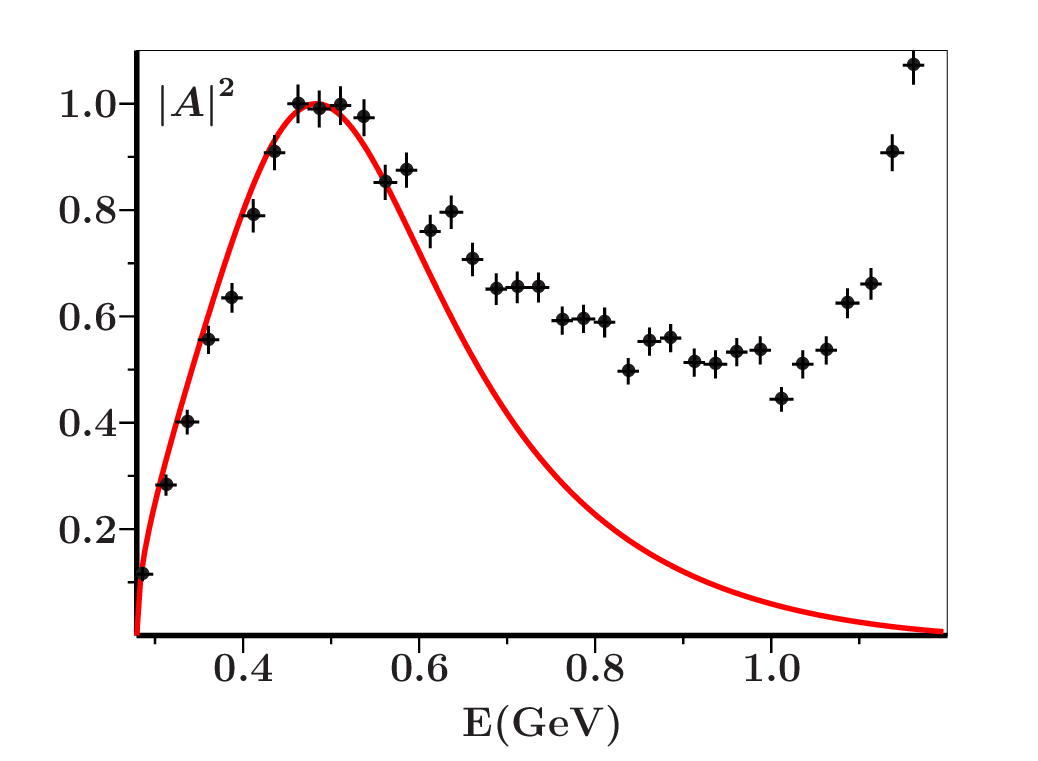}}\\ [-150pt]
\hspace{100pt}(a) & \hspace{100pt}(b)\\ [120pt]
\end{tabular}
\end{center}
\caption[]{\small The RSE $S$-wave $I=0$ $\pi\pi$ distribution
(arbitrary units), corrected for phase space (red curves),
compared to the $\pi\pi$ $S$-wave signal of
Ref.~\cite{PLB598p149} extracted from $J/\psi\to\omega\pi\pi$.
In (a) we compare the RSE to the $\sigma$ signal extracted
from the data by BES (blue histogram), while in
(b) it is compared to the full BES $\pi\pi$ signal.}
\label{BesFigures}
\end{figure}

Hence, for the $S$-wave meson-meson distributions in decay processes
$M\to\xrm{spectator}+\xrm{meson}+\xrm{meson}$, assuming that indeed
final-state interactions with the spectator can be neglected, we obtain
\begin{equation}
A_{0}^{0}(E)\;\propto\;
\lambda\,
\fndrs{2pt}{\sin (pa)}{-2pt}{pa}\,
\left[
1-2\lambda^{2}\,\mu\,
\fndrs{2pt}{\sin (pa)}{-2pt}{pa}\, e^\x{ipa}\,
\dissum{n=0}{\infty}\,\fndrs{3pt}{(n+1)\, 4^{-n}}{-3pt}{E_{n}-E}
\right]^{-1}
\;\;\; .
\label{ARSE}
\end{equation}
In Fig.~\ref{BesFigures} we compare
the $I=0$ $S$-wave $\pi\pi$ amplitudes
resulting from expression (\ref{ARSE}),
with the amplitudes observed
by the BES collaboration \cite{PLB598p149}.
The parameters for the theoretical curves of Figs.~\ref{BesFigures}
are identical to the parameters employed
for the theoretical result of Fig.~\ref{t00Figures}.
The experimental data shown in Fig.~\ref{BesFigures}a
are identified by the BES collaboration \cite{PRL96p132001}
as the resonant $I=0$ $S$-wave component, i.e., the $\sigma$ meson.
The agreement between the RSE and the BES $\sigma$ data is reasonable, but
not very good, at least for the parameter set used to reproduce the
elastic $I=0$ $S$-wave $\pi^{+}\pi^{-}$ data of Ref.~\cite{PRL96p132001}.
For low energies, say below 600 MeV, this slight discrepancy can be
understood from the lower-lying $\sigma$ pole and somewhat larger  $I=0$
scattering length in the CCL parametrisation, as compared to the BES analysis
\cite{JPG34p151}. Another reason, relevant at higher energies, is that the
RSE does not isolate resonant parts from so-called non-resonant parts of the
scattering data.  Hence, instead of forcing the RSE to agree with the $\sigma$
data by changing the parameters, we better compare the RSE to the full data.
In Fig.~\ref{BesFigures}b a comparison is shown of the RSE formula (\ref{ARSE})
with the full $\pi\pi$ data, obtained in Ref.~\cite{PLB598p149} from
$J/\psi\to\omega\pi\pi$.

A few words of caution are in place here.
As already indicated above, the BES collaboration found a higher-lying sigma
pole in their analysis than CCL, namely at $(541\pm 39)-i(252\pm 42)$ MeV
\cite{PLB598p149}, to be compared to the reported \cite{PRL96p132001} CCL
value of $(441\pm 4)-i(272\pm 6)$ MeV.
Hence, the theoretical curve of Fig.~\ref{BesFigures}a
should \em a priori \em \/not follow the BES data very closely.
Nevertheless, our prediction also seems to agree quite well, up to about
600 MeV, with the full BES $\pi\pi$ signal in Fig.~\ref{BesFigures}b.
However, BES reported contamination of the data from processes of a different
origin, so that no definite conclusions on pole positions and line shapes
can be drawn from this comparison. For all that, it seems safe to conclude
that the RSE is in fact capable of reasonably describing both the elastic
scattering data of Ref.~\cite{PRL96p132001} and the production data of BES
\cite{PLB598p149}, with the same set of parameters, contrary to what was
suggested by C.~Hanhart in Ref.~\cite{HEPPH0609136}.

We observe that the RSE prediction follows
the data \cite{PLB598p149} closely up to about 0.6 GeV, which is the limit
of validity of the approximation (\ref{Tfull}), which neglects $s\bar{s}$ and
$K\bar{K}$ contributions. The main differences between the RSE and the BES
data stem from the $D$-wave $f_{2}(1270)$ resonance, and also from a small
signal due to the $f_{0}(980)$ resonance, both not contemplated in formula
(\ref{ARSE}). It should furthermore be noticed that the pole structure of the
amplitude in Eq.~(\ref{ARSE}) is exactly the same as of the $S$-wave term in
the elastic scattering amplitude (\ref{Tfull}). For the parameter set given
in the caption of Fig.~\ref{t00Figures}, we find the $\sigma$ pole at
$462-i207$ MeV.

\begin{figure}[htbp]
\begin{center}
\begin{tabular}{cc}
\resizebox{0.42\textwidth}{!}{\includegraphics{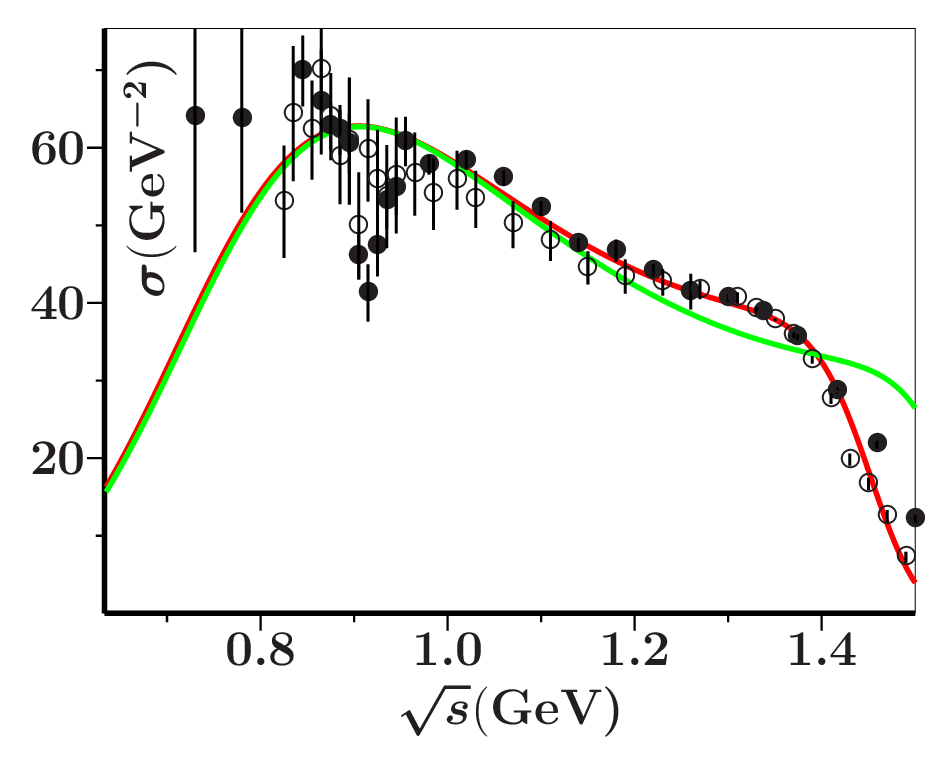}} &
\resizebox{0.47\textwidth}{!}{\includegraphics{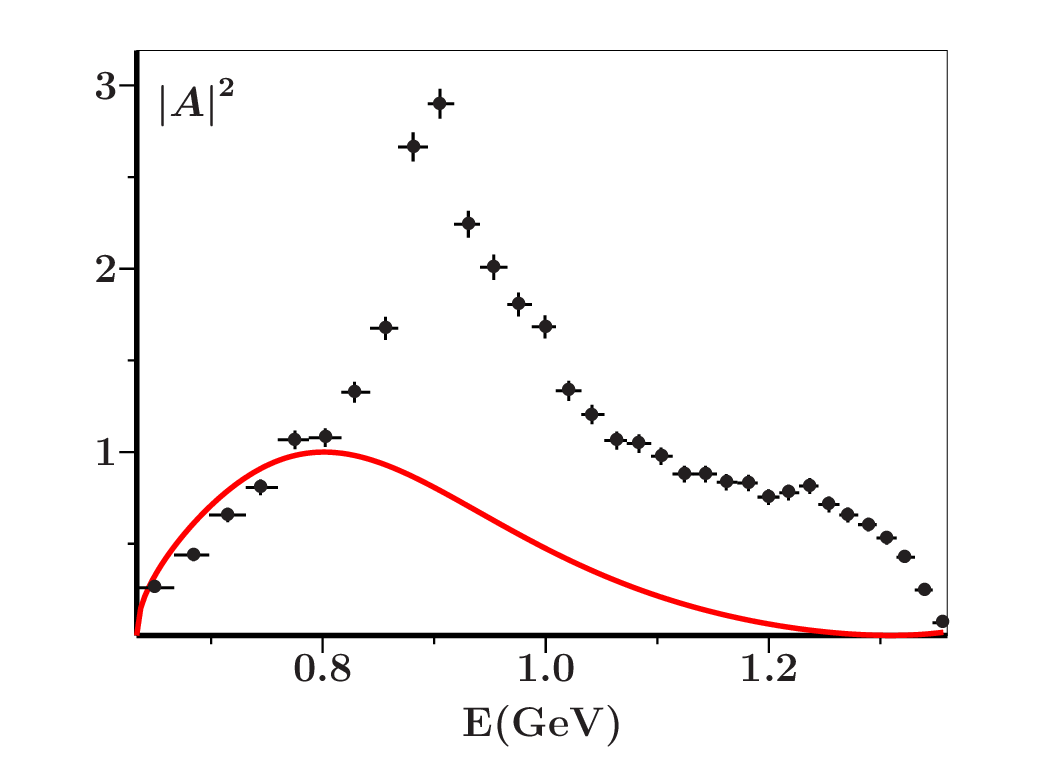}}\\ [-150pt]
\hspace{170pt}(a) & \hspace{168pt}(b)\\ [120pt]
\end{tabular}
\end{center}
\caption[]{\small The $I=1/2$ $S$-wave $K\pi$ distribution,
for elastic scattering (a) and production (b, arbitrary units).
In (a) the prediction of Eq.~(\ref{Tfull}) (red curves) is compared to
the available scattering data \cite{NPB133p490} ($\odot$) and
\cite{NPB296p493} ($\bullet$). In (b) the result of Eq.~(\ref{ARSE})
(red curve) is compared to the E791 analysis of
$D^{+}\to K^{-}\pi^{+}\pi^{+}$ \cite{PRL89p121801}.
The green curve of (a) is used in section~\ref{Pwavestudy}.}
\label{KpiFigures}
\end{figure}

Turning now to the kaon-pion system, in Fig.~\ref{KpiFigures} we compare
the $I=1/2$ $S$-wave $K\pi$ component of the elastic-scattering
formula~(\ref{Tfull}) with the data of E.~Estabrooks {\it et al.}
\cite{NPB133p490} as well as the LASS data \cite{NPB296p493},
and the same component of the production formula~(\ref{ARSE}) with the
$D^{+}\to\pi^{+} K^{-}\pi^{+}$ data of the E791 collaboration
\cite{PRL89p121801}.
We obtain good agreement between the RSE and data, apart from a signal
which is mainly due to the $P$-wave resonance $K^{\ast}(892)$.
We find a contribution of about 41\% from the $K_{0}^{\ast}(800)$
(alias kappa) resonance to the total signal. The $\kappa$ pole comes out
at $772-i281$ MeV \cite{AIPCP814p143}.

Now, although the RSE amplitude (\ref{ToVfull}) seems to agree well
with the BES (and also the E791) data, it should be noticed that the
experimental signal stems from a three-body decay process, whereas the
theoretical curve in Fig.~\ref{BesFigures}b concerns a two-body amplitude.
So it is not obvious whether the comparison makes sense at all.
In the following, we shall study this issue in a bit more detail, viz.\ for
the 3-body decay processes
\begin{equation}
D^{+}\to K^{+}\pi^{+}\pi^{-}
\;\;\;\;\xrm{and}\;\;\;\;
D^{+}_{s}\to K^{+}\pi^{+}\pi^{-}
\;\;\; .
\label{DXtoKpipi}
\end{equation}

\begin{figure}[htbp]
\begin{center}
\begin{tabular}{cc}
\resizebox{0.47\textwidth}{!}{\includegraphics{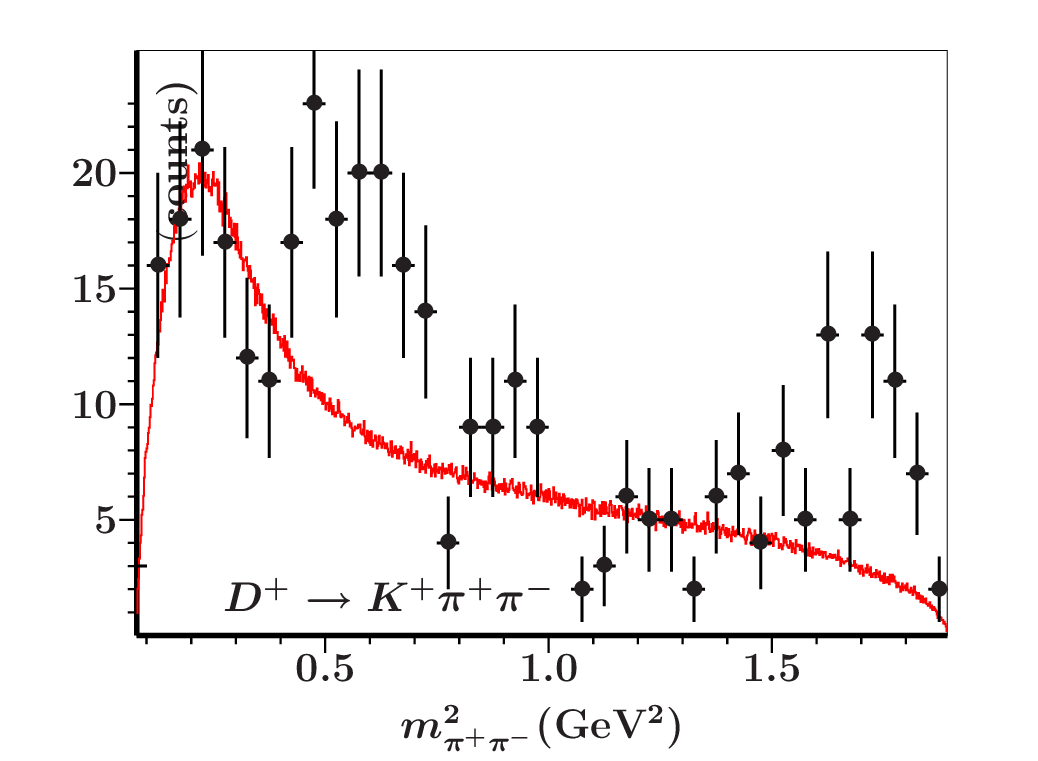}} &
\resizebox{0.47\textwidth}{!}{\includegraphics{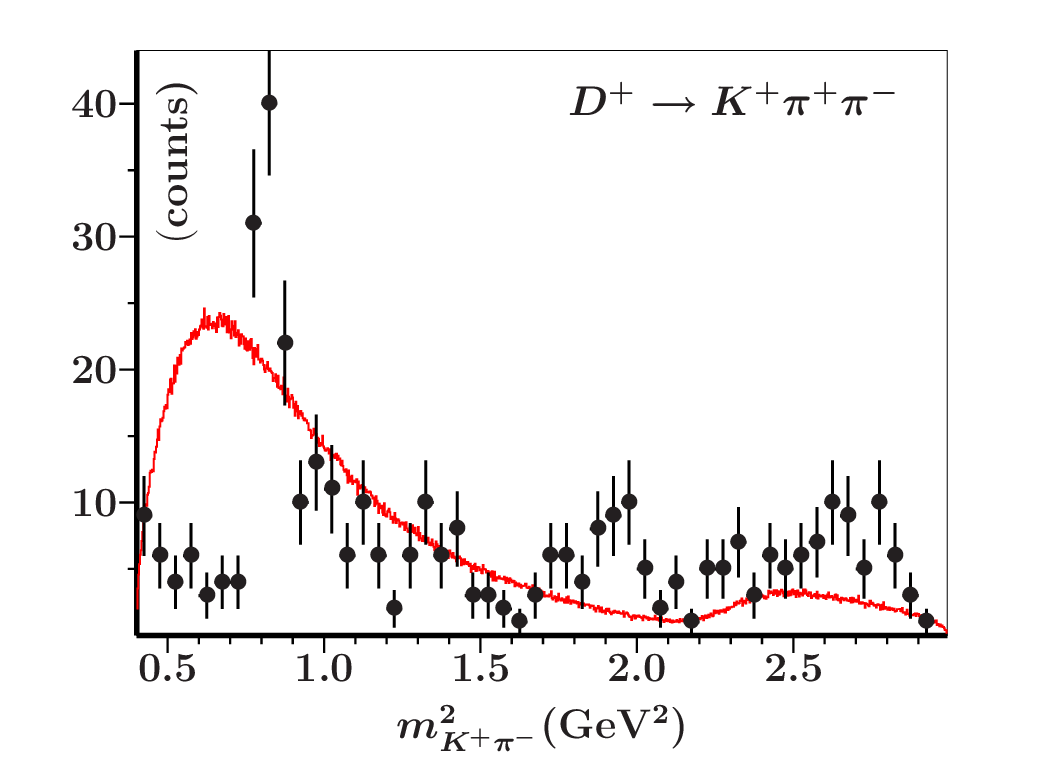}}\\ [10pt]
\resizebox{0.47\textwidth}{!}{\includegraphics{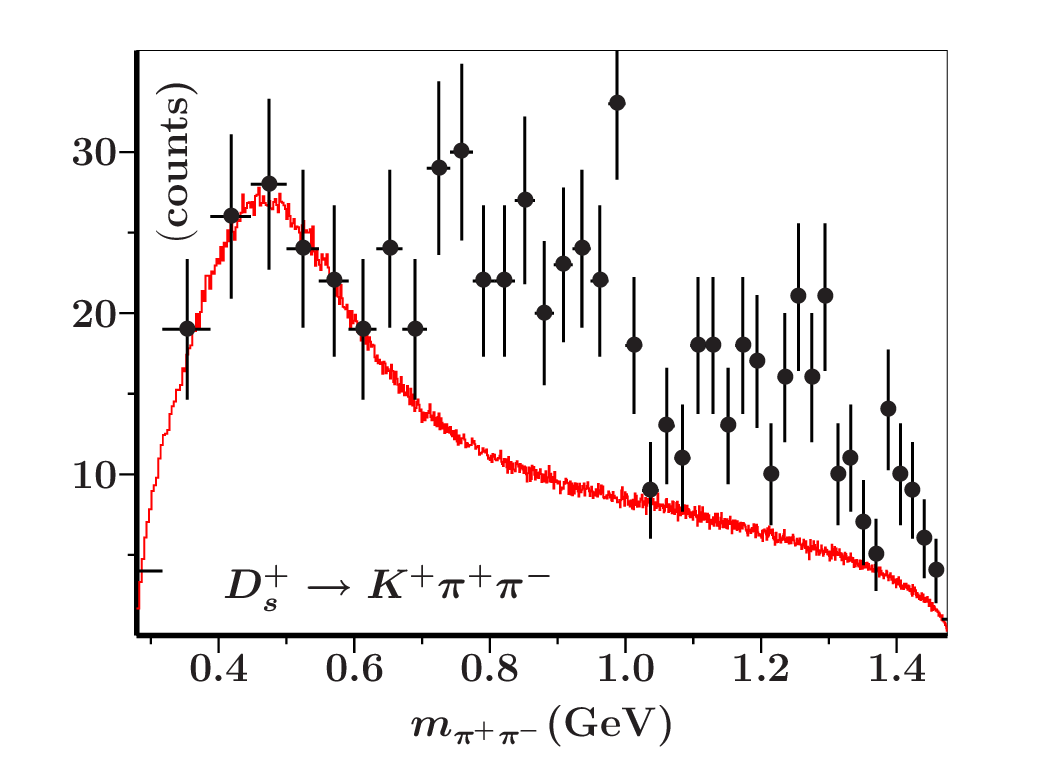}} &
\resizebox{0.47\textwidth}{!}{\includegraphics{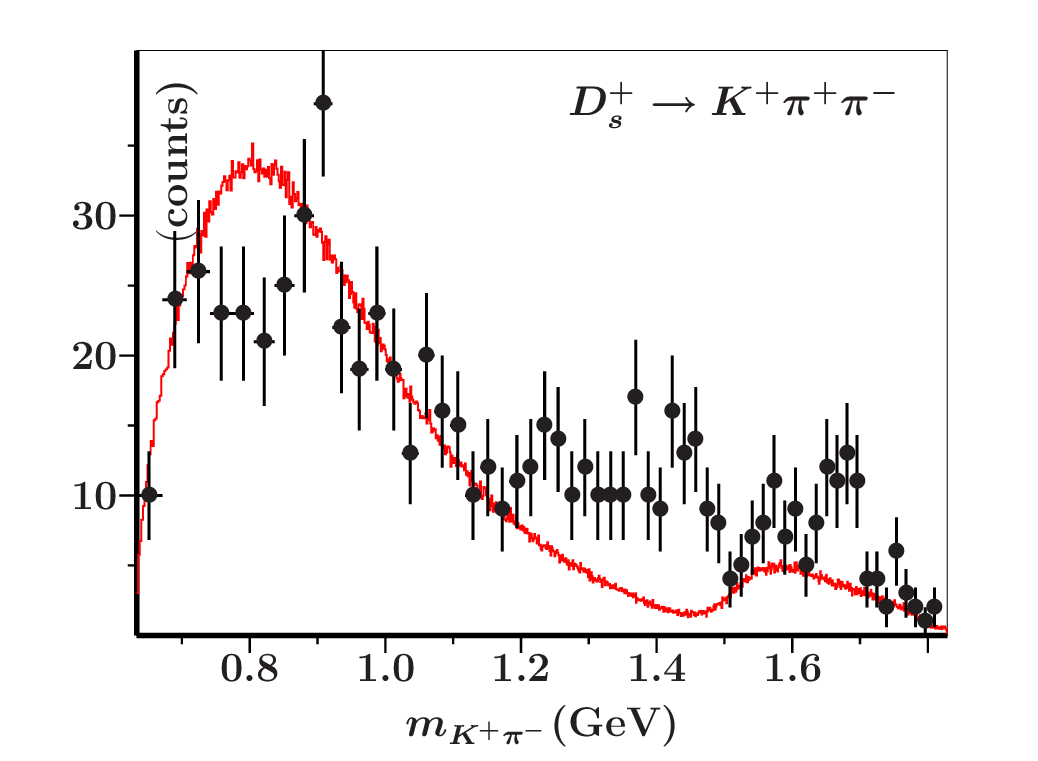}}\\ [-20pt]
\end{tabular}
\end{center}
\caption[]{\small Comparison of the Dalitz-plot projections
for the RSE two-particle $S$-wave contributions
to $D^{+}\to K^{+}\pi^{+}\pi^{-}$
and $D^{+}_{s}\to K^{+}\pi^{+}\pi^{-}$,
and the results of the FOCUS experiment \cite{PLB601p10}.}
\label{FocusDps}
\end{figure}

Here, we assume that, after the initial flavour content of the charmed mesons
has been transformed by weak processes into $u\bar{s}$, the three-body decays
under consideration take place as a cascade of two OZI-allowed \cite{OZI}
two-body processes. As a consequence, one of the three decay products may be
considered a spectator. In this approximation, we may write the total
amplitude in the form
\begin{eqnarray}
\lefteqn{A_\xrm{\scriptsize tot}=}
\label{Atot}\\ [10pt] & &
A\left(\left[ I=0,\; S\;\xrm{wave}\right]\, K^{+}\right) +
A\left(\left[ I=1/2,\; S\;\xrm{wave}\right]\,\pi^{+}\right)\,
+\xrm{higher waves and different isospins}
\; .
\nonumber
\end{eqnarray}
At first, we shall neglect the higher waves, as well as possible different
isospin combinations, and take for the total amplitude of the above decay
processes just the sum of the first two terms in Eq.~(\ref{Atot}),
i.e., terms of the form (\ref{ToVfull}).
In order to evaluate such an expression, we generate $10^{6}$ events for the
allowed three-body phase space.
The resulting Dalitz plot \cite{PM44p1068} is projected onto
the $K^{+}\pi^{-}$ and $\pi^{+}\pi^{-}$ 2-body channels.
In Fig.~\ref{FocusDps} we
compare the RSE results to the FOCUS data \cite{PLB601p10}
for the decay processes given in Eq.~(\ref{DXtoKpipi}).
\clearpage

One difference between the RSE signals shown
in Figs.~\ref{BesFigures}, \ref{KpiFigures} on the one hand,
and in Fig.~\ref{FocusDps} on the other hand,
is striking, namely that in the latter (production) case the amplitudes
do not vanish anywhere except for the boundaries of allowed phase space.
The reason appears to be simple. Namely, each of the two $S$-wave
amplitudes of formula (\ref{Atot}) has its own zeros associated with it,
which lie at the energies of the corresponding confinement spectra, and are
different for different flavour combinations (see formula (\ref{t00HO})).
Hence, the sum will in general not vanish, particularly not for a Dalitz-plot
projection at a constant invariant mass in one of the two $S$-wave amplitudes,
integrated over all phase-space-allowed invariant masses for the other
$S$-wave amplitude.

The data have been collected in equal bins of invariant mass squared
by the FOCUS collaboration, and so are the RSE results.
However, we prefer to present the results as a function of mass,
rather than mass squared. Instead of manipulating the data for that purpose,
we let the bin size vary with mass in Fig.~\ref{FocusDps}.
The RSE results for the processes (\ref{DXtoKpipi})
have been normalised by hand, in order to be comparable
to the FOCUS data \cite{PLB601p10}.
Our choice of normalisation suggests that the dominant modes are given by
two-body $S$-waves. The remaining structures can visibly be associated with
the $f_{0}(980)$ (dominantly $s\bar{s}$) component of isoscalar $0^{++}$
states (here, we only consider the $n\bar{n}$ component),
as well as $P$- and $D$-wave contributions,
like the $\rho (770)$ and the $f_{2}(1270)$ in $\pi^{+}\pi^{-}$,
and the $K^{\ast}(892)$, $K^{\ast}(1410)$, $K^{\ast}_{2}(1430)$ and
$K^{\ast}(1680)$ in $K^{+}\pi^{-}$, besides their interference patterns with
the $S$-wave signals.

The RSE amplitude (\ref{Atot}), which consists of the sum of two amplitudes
of the form (\ref{ARSE}), has the same pole structure as the individual
$S$-wave isoscalar $\pi\pi$ and isodoublet $K\pi$ elastic-scattering
amplitudes. So one can only determine the pole positions in the production
amplitude once the model parameters have been fixed. Moreover, one cannot
just remove a pole from this amplitude, in order to verify whether or not
it contributes significantly to the experimental signal.
Whereas, in most cases, one can relate resonant structures in elastic
scattering to each of the poles in the scattering amplitude, here such a
relation is not evident at all, since most of the poles \cite{AIPCP814p143}
in the $\pi\pi$ and $K\pi$ $S$-wave amplitudes do not show up as peaked
structures in Fig.~\ref{FocusDps} for the total RSE amplitude (\ref{Atot}).
In fact, most of the theoretical $S$-wave signal has the aspect
of a flat non-resonating background, except for the clear $\sigma$ and
$\kappa$ signals at 470 MeV in $\pi^{+}\pi^{-}$ and 800 MeV in
$K^{+}\pi^{-}$, respectively. Nevertheless, the amplitudes (\ref{ARSE})
contain {\it all}\/ poles of elastic $S$-wave scattering. Furthermore, it is
not easy to find out whether the theoretical structure in $K\pi$
around 1.6 GeV stems from the lower pole, which corresponds to the
$K_{0}(1430)$ resonance in elastic scattering, or from a higher pole.

\section{Study of the \bm{P} wave contributions}
\label{Pwavestudy}

Although the RSE results shown in Fig.~\ref{FocusDps}
agree well with experiment \cite{PLB601p10},
it must be acknowledged that in the $K\pi$ Dalitz-plot projections
at low energies, which is precisely the mass region where the RSE is
supposed to work well, the predicted amplitudes for
$D^{+}_{s}\to K^{+}\pi^{+}\pi^{-}$ and $D^{+}\to K^{+}\pi^{+}\pi^{-}$
are larger respectively much larger than those reported in
Ref.~\cite{PLB601p10}.
Hence, the legitime question arises whether the interferences of the
$\pi\pi$ and $K\pi$ $S$-waves with higher waves
are capable of correcting these discrepancies,
as claimed in the previous section.
Thereto, we embark on an extension of expression (\ref{Atot})
to $P$-waves, i.e.,
\begin{eqnarray}
\lefteqn{A_\xrm{\scriptsize tot}\, =\,
A\left(\left[ I=0,\; S\;\xrm{wave}\right]\, K^{+}\right)
\, +\, A\left(\left[ I=1/2,\; S\;\xrm{wave}\right]\,\pi^{+}\right)
\, +}
\label{AtotSP}\\ [10pt] & & \;\;\;\;\;\;\;\;\;
+\, A\left(\left[ I=0,\; P\;\xrm{wave}\right]\, K^{+}\right)
\, +\, A\left(\left[ I=1/2,\; P\;\xrm{wave}\right]\,\pi^{+}\right)
\; .
\nonumber
\end{eqnarray}
The generalisation \cite{IJTPGTNO11p179} of the RSE two-particle
one-channel amplitude (\ref{ARSE}) to higher waves is given by
\begin{equation}
A_{\ell}(E)\;\propto\;
\lambda_{q\bar{q}}\,
i^{\ell}\, j_{\ell}\left( pa_{q\bar{q}}\right)\,
\left[
1-2i\lambda_{q\bar{q}}^{2}\,\mu\,
pa_{q\bar{q}}\,
j_{\ell}\left( pa_{q\bar{q}}\right)\,
h^{(1)}_{\ell}\left( pa_{q\bar{q}}\right)\,
\dissum{n=0}{\infty}\,\fndrs{5pt}{g^{2}_{n\ell}}{-3pt}{E_{n}-E}
\right]^{-1}
\;\;\; ,
\label{AellRSE}
\end{equation}
where \cite{ZPC21p291}
$g^{2}_{n0}=(n+1)4^{-n}$ and $g^{2}_{n1}=\frac{1}{3}(2n+3)4^{-n}$,
and where $a_{q\bar{q}}$ and $\lambda_{q\bar{q}}$
can be related to the flavour-independent
$^{3}P_{0}$ coupling $\lambda$ and average transition distance $a$ by
\begin{equation}
\left(\begin{array}{c} \lambda_{q\bar{q}}\\ [10pt] a_{q\bar{q}}
\end{array}\right)\, =\,
\sqrt{\fnd{\omega\left( m_{q}+m_{\bar{q}}\right)}{m_{q}m_{\bar{q}}}}\,
\left(\begin{array}{c} \lambda\\ [10pt] a\end{array}\right)
\;\;\; .
\label{lambdaRSE}
\end{equation}

The procedure to calculate the total amplitude for each of
the possible invariant-mass combinations of $\pi\pi$ and $K\pi$,
is as follows.
First, we adjust the only ``free'' parameters, viz.\
$\lambda$ and $a$, of each of the two-particle $P$ waves
to the elastic scattering data,
such that the $\rho$ and $K^{\ast}(892)$ resonances
are well reproduced.
Then, we adjust the fraction with which each of the two $P$-wave
amplitudes contributes to the total amplitude (\ref{AtotSP}).
Apart from an overall phase of $90^{\circ}$ between the $P$- and $S$-waves,
there is no need to introduce additional relative phases.
Hence, in total, the relative contributions of the four terms
in expression (\ref{AtotSP}) can be parametrised by 3 real numbers.
\begin{figure}[htbp]
\begin{center}
\begin{tabular}{cc}
\resizebox{0.47\textwidth}{!}{\includegraphics{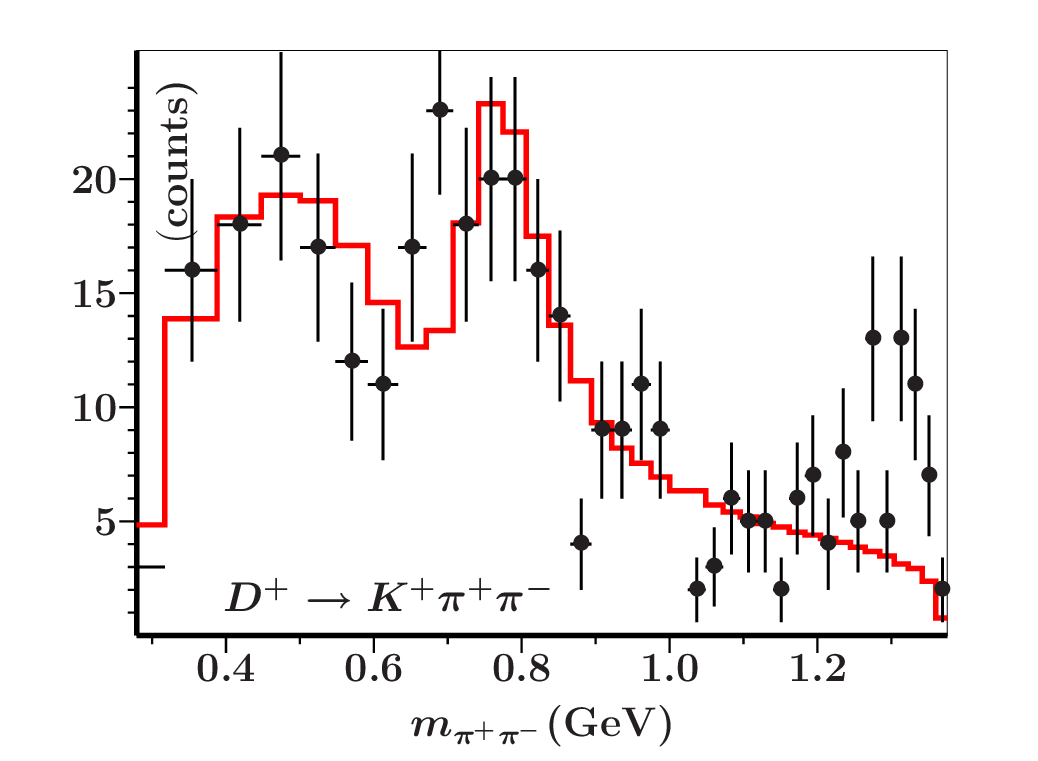}} &
\resizebox{0.47\textwidth}{!}{\includegraphics{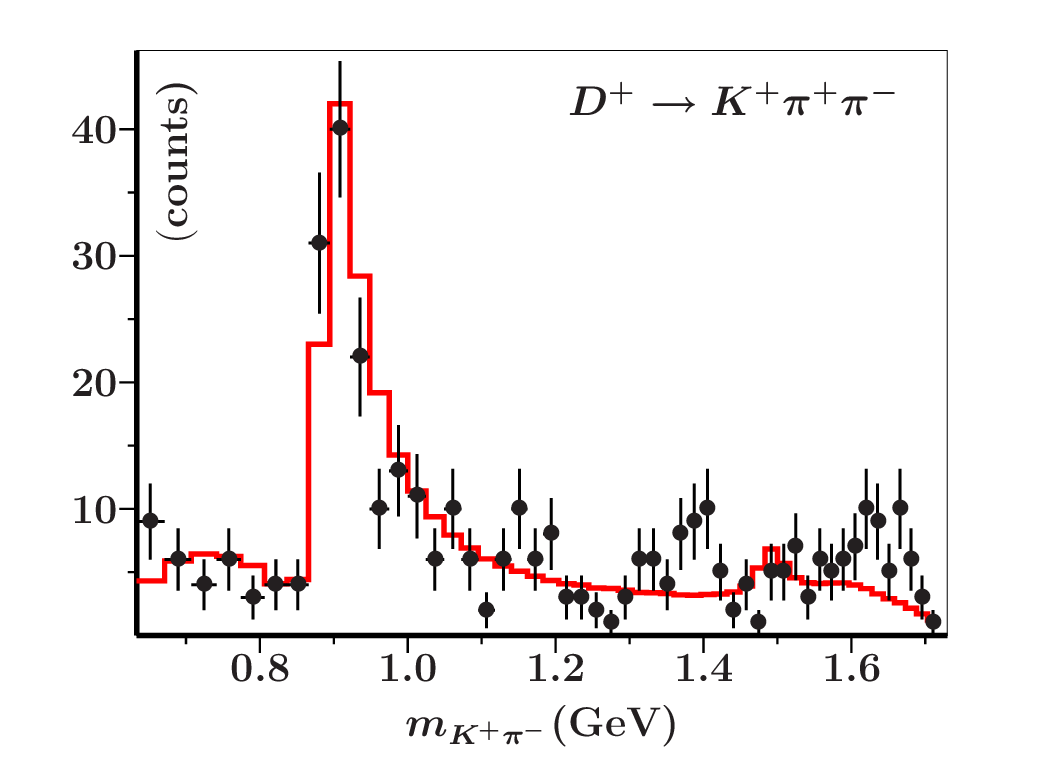}}\\ [10pt]
\resizebox{0.47\textwidth}{!}{\includegraphics{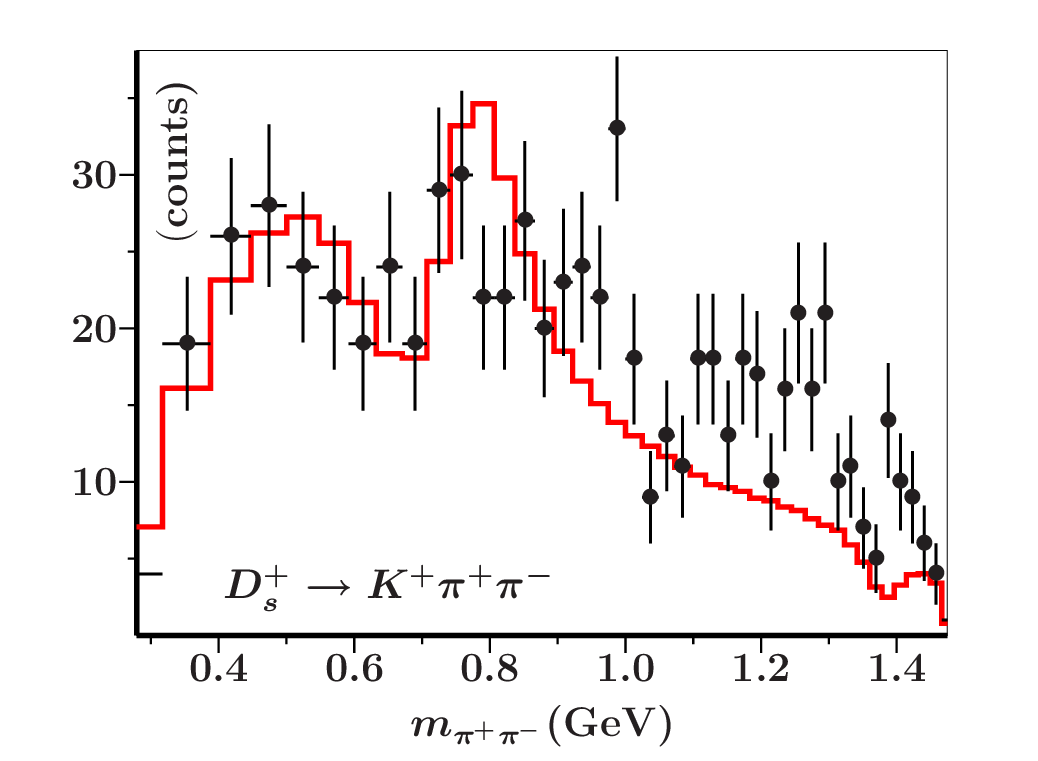}} &
\resizebox{0.47\textwidth}{!}{\includegraphics{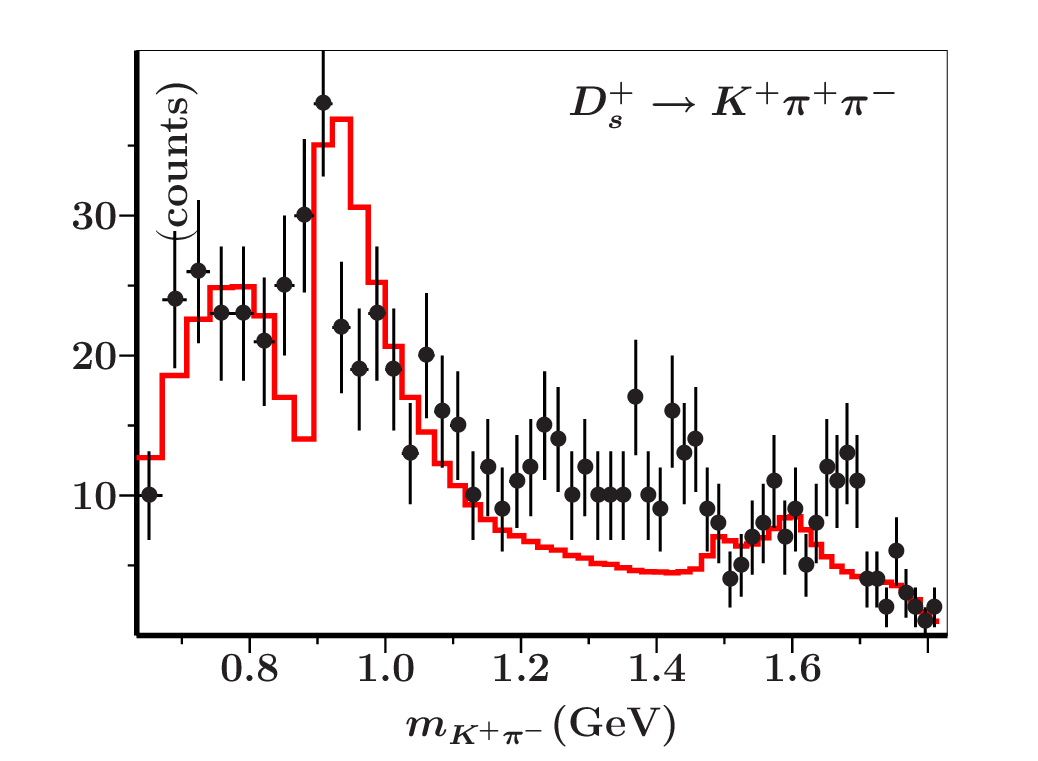}}\\ [-20pt]
\end{tabular}
\end{center}
\caption[]{\small Comparison of the Dalitz-plot projections
for the RSE two-particle $S$- and $P$-wave contributions
to $D^{+}\to K^{+}\pi^{+}\pi^{-}$
and $D^{+}_{s}\to K^{+}\pi^{+}\pi^{-}$,
and the results of the FOCUS experiment \cite{PLB601p10}.
The total number of events under the theoretical curves is 10 percent less
than in experiment for $D^{+}$ decay, and 20 percent for $D^{+}_{s}$,
in order to account for the signal which is missing in the theoretical
curves at higher energies.}
\label{FocusDpsSP}
\end{figure}

It is clear from the resulting amplitudes
shown in Fig.~\ref{FocusDpsSP},
that the interference between $S$- and $P$-waves can perfectly
account for the discrepancies observed in Fig.~\ref{FocusDps}.
In particular, the $\kappa$ resonance seems to have disappeared from
the $D^{+}\to K^{+}\pi^{+}\pi^{-}$ signal,
in agreement with experiment.
Nevertheless, the RSE amplitude for $D^{+}\to K^{+}\pi^{+}\pi^{-}$
does contain the $\kappa$ pole, namely at $758-i279$ MeV.
Consequently, its apparent absense from the signal does not mean that it
does not contribute to the amplitude.

In Table~\ref{relativeintensities} we show the relative intensities
of the four contributions of Eq.~(\ref{AtotSP}).
\begin{table}[htbp]
\begin{center}
\begin{tabular}{||r|c|c|c||}
\hline\hline & & & \\ [-5pt]
\begin{tabular}{c} relative\\ratios\end{tabular} &
$\fndrs{5pt}{\left( K^{+}\pi^{-}\right)_{S}\pi^{+}}
{-5pt}{\left(\pi^{+}\pi^{-}\right)_{S}K^{+}}$ &
$\fndrs{5pt}{\left(\pi^{+}\pi^{-}\right)_{P}K^{+}}
{-5pt}{\left(\pi^{+}\pi^{-}\right)_{S}K^{+}}$ &
$\fndrs{5pt}{\left( K^{+}\pi^{-}\right)_{P}\pi^{+}}
{-5pt}{\left(\pi^{+}\pi^{-}\right)_{S}K^{+}}$\\ [15pt]
\hline & & & \\ [-5pt]
for $D^{+}$ decay & 1.5 & 1.7 & 1.4\\ [10pt]
for $D^{+}_{s}$ decay & 2.5 & 2.0 & 0.7 \\ [10pt]
\hline\hline
\end{tabular}
\end{center}
\caption[]{\small
The relative intensities of the four amplitudes of
Eq.~(\ref{AtotSP}).
}
\label{relativeintensities}
\end{table}

\begin{figure}[htbp]
\begin{center}
\begin{tabular}{c}
\resizebox{0.47\textwidth}{!}{\includegraphics{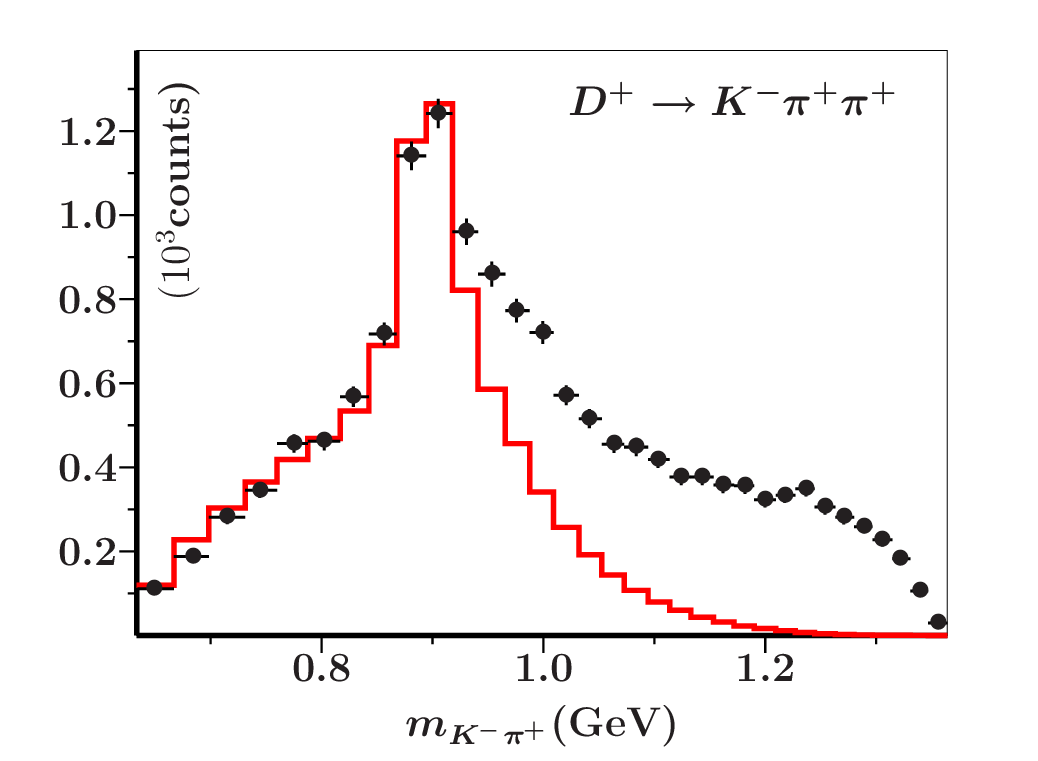}}
\end{tabular}
\end{center}
\caption[]{\small Comparison of the Dalitz-plot projections
for the RSE two-particle $S$- and $P$-wave contributions
to $D^{+}\to K^{-}\pi^{+}\pi^{+}$,
and the results of the E791 experiment \cite{PRL89p121801}.
The relative intensity of the $P$-wave contribution is 0.35
with respect to the $S$-wave contribution,
and has a relative phase of $180^{\circ}$ for the theoretical curve.
The RSE parameters are taken from Table~\ref{modelparameters}.}
\label{KpifiguresSP}
\end{figure}

In Fig.~\ref{KpifiguresSP}, we also find improvement at low energies
with respect to Fig.~\ref{KpiFigures}b, by including the $P$-wave amplitude
for the process $D^{+}\to K^{-}\pi^{+}\pi^{+}$, measured by the E791
collaboration \cite{PRL89p121801}.
The data of Fig.~\ref{KpifiguresSP} were collected by E791 by
selecting those $K^{-}\pi^{+}$ invariant masses which are lowest in mass
for the two possible $K^{-}\pi^{+}$ combinations in the decay process
$D^{+}\to K^{-}\pi^{+}\pi^{+}$.
In our Monte-Carlo event generation we perform the same selection procedure
for the RSE.

We recall that the production processes
$D^{+}\to K^{+}\pi^{+}\pi^{-}$
and
$D^{+}\to K^{-}\pi^{+}\pi^{+}$
are different in nature.
This difference reflects itself, in particular,
in different relative phases, viz.\
$90^{\circ}$ for the former process and $180^{\circ}$ for the latter.
Hence, the line shapes of Figs.~\ref{FocusDpsSP} and \ref{KpifiguresSP}
are very different.
Nevertheless, as in the $D^{+}\to K^{+}\pi^{+}\pi^{-}$ cross section,
also in the $D^{+}\to K^{-}\pi^{+}\pi^{+}$ signal
there is no clear presence of the $\kappa$ enhancement,
although the RSE amplitude contains the $\kappa$ pole
at $758-i279$ MeV.

\section{Further discussion of the RSE}
\label{RSE2}

Formulae (\ref{ARSE}) and (\ref{AellRSE}) are very incomplete expressions for
the amplitudes of meson-meson production. First of all, they are based on a
set of dynamical equations in which the interaction describing the transition
from quark-antiquark to meson-meson, via the $^{3}P_{0}$ mechanism, is modelled
through a one-delta-shell potential with radius $a$. The advantage of such an
approach is that the confinement spectrum~(\ref{t00HO}) enters the expressions
explicitly. However, it has the consequence that $a$ must be adjusted
for each separate configuration, thus losing universality. Secondly, by
reducing to one the number of possible channels to which a $q\bar{q}$ state is
allowed to couple, one must adjust also $\lambda$, in order to get the
elastic-scattering poles in the correct positions. So one also loses the
universality of the $^{3\!}P_{0}$ coupling. On the other hand, the expressions
for the amplitudes become much less transparent when all ingredients of flavour
independence and channel completeness are built in. In the present paper, our
principal aim is to show that the RSE reproduces the basic properties
of meson-meson production, and not to carry out a detailed data analysis.
The price we have to pay is that $a$ and $\lambda$ are not completely
constant for the four amplitudes in question. In Table~\ref{modelparameters} we
give the values used for the curves in Fig.~\ref{FocusDpsSP}.
\begin{table}[htbp]
\begin{center}
\begin{tabular}{||r|c|c|c|c||}
\hline\hline & & & & \\ [-5pt]
\begin{tabular}{c} model\\ parameters\end{tabular} &
$\left(\pi^{+}\pi^{-}\right)_{S}K^{+}$ &
$\left( K^{+}\pi^{-}\right)_{S}\pi^{+}$ &
$\left(\pi^{+}\pi^{-}\right)_{P}K^{+}$ &
$\left( K^{+}\pi^{-}\right)_{P}\pi^{+}$\\ [15pt]
\hline & & & & \\ [-5pt]
$\frac{1}{3}a$ & 1.00 & 1.09 & 1.12 & 1.01\\ [10pt]
$\frac{3}{4}\lambda$ & 1.000 & 0.654 & 0.960 & 0.995\\ [10pt]
\hline\hline
\end{tabular}
\end{center}
\caption[]{\small
The values of $a$ and $\lambda$ in each of the four amplitudes
of Eq.~(\ref{AtotSP}).
}
\label{modelparameters}
\end{table}

The flavour-independent coupling constant $\lambda$ represents
the intensity of $q\bar{q}$ pair creation and annihilation,
which is considered a constant in the RSE in case all meson-meson channels
allowed by quantum numbers are taken into account. In practice, the latter
is an impossible task, of course. However, omitting meson-meson channels
has an influence on the resulting resonance pole positions. Thus, the
coupling constants of the remaining channels act as effective parameters,
which must be adjusted in order to reproduce the elastic scattering data.
At low energies, there is a negligible effect on pole positions
and resonance shapes from  meson-meson channels with much higher thresholds.
Accordingly, their omission has little influence on $\lambda$. So, for
energies below 1 GeV, it is a reasonable approximation to restrict the
allowed meson-meson channels to those containing a pair of pseudoscalar
mesons, and even to only $\pi\pi$ and $K\pi$, for the isoscalar and
isodoublet cases, respectively. The only situation where this philosophy
does not work well here is for the $\kappa$(800) pole. The reason is that
the $K\eta '$ channel takes \cite{PLB454p165} about half of the total
coupling, but nevertheless does not contribute significantly to the $\kappa$
pole position, since this high-lying channel is strongly damped there.
As a consequence, in this case the effective coupling for $K\pi$ to
$J^{P}=0^{+}$ $n\bar{s}$ comes out some 35 percent below the average
flavour-independent value of $\lambda$.

The effective average interaction radius $a$ mainly controls the resonance
width, which is relatively sensitive to the omission of higher thresholds,
as can be observed from Table~\ref{modelparameters}.

The effective flavour masses $m_{u/d}=406$ MeV and $m_{s}=508$ MeV,
as well as the average mesonic level splitting $\omega =0.19$ GeV,
have been determined in Ref.~\cite{PRD27p1527}.
These parameters generate the quenched quark-antiquark spectrum,
which should be compared to corresponding lattice calculations.
For example, in a very recent preprint \cite{HEPLAT0702023} the Scalar
Collaboration found 1.58$\pm$0.35 GeV for the lightest isoscalar scalar meson
and 1.71$\pm$0.54 GeV for its isodoublet partner,
which in the RSE (Ref.~\cite{ZPC30p615}, 1986)
came out (Eq.~(\ref{t00HO}) for $E_{01}=m_{q}+m_{\bar{q}}+5\omega /2$)
at 1.29 and 1.39 GeV, respectively. However, by assuming the latter values
in the quenched approximation, one can in the RSE even determine the lowest
$S$-matrix poles \cite{ZPC30p615},
by adjusting the model parameters $a$ and $\lambda$
to the available elastic $S$- \cite{ZPC30p615} and $P$-wave \cite{PRD27p1527}
scattering data.
Moreover, as mentioned above, the very string-breaking
picture underlying the RSE, and even the empirically found values for $a$
have received strong support from recent unquenched lattice calculations
\cite{PRD71p114513}.
Therefore, the RSE is really capable of relating experiment to QCD,
via the lattice.

In previous work \cite{AIPCP814p143}, and also for the theoretical curves
in red shown in Fig.~\ref{KpiFigures},
we used a value of 1.31 GeV for the ground state of
the $J^{P}=0^{+}$ $n\bar{s}$ confinement spectrum, which is about 80 MeV below
its value following from Eq.~(\ref{t00HO}), in order to account for the
contribution of the $K\eta '$ channel. Here, we have restored this, leading to
small adjustments for $a$ and $\lambda$, and to the green curve in
Fig.~\ref{KpiFigures} for the elastic RSE amplitude.
The corresponding $\kappa$ pole thus moves from \cite{AIPCP814p143}
$772-i281$ MeV to $758-i279$ MeV.

The ``poles'' of the confinement spectrum, which enter formula
(\ref{AellRSE}), are then fully determined by the parameters
of Ref.~\cite{PRD27p1527}, i.e.,
$E_{n\ell_{q\bar{q}}}=m_{q}+m_{\bar{q}}+(2n+\ell_{q\bar{q}}+1.5)\omega$.
This amounts, for the $S$-wave amplitudes of Eq.~(\ref{AtotSP})
($\ell_{q\bar{q}}=1$), to 1.29, 1.67, 2.05, \ldots GeV for the isoscalars,
and 1.39, 1.77, 2.15, \ldots GeV for the isodoublets.
Notice that the $\kappa$ pole lies about 600 MeV below the lowest
``pole'' of the input confinement spectrum
(see also Fig.~1 of Ref.~\cite{HEPPH0703256}).
The reason is that resonance poles do not occur for $E=E_{n}$
in formula (\ref{AellRSE}), but only when the denominator vanishes.
In Ref.~\cite{EPJC22p493} pole trajectories were shown
for variations in the parameter $\lambda$ of formula (\ref{AellRSE}).

For the $P$-wave amplitudes of Eq.~(\ref{AtotSP}) ($\ell_{q\bar{q}}=0$),
we have a spectrum at 1.10, 1.48, 1.86, \ldots GeV for the isoscalars,
and 1.20, 1.58, 1.96, \dots GeV for the isodoublets.
Furthermore, for the parameter set of Table~\ref{modelparameters},
we find the $\sigma$ pole at $462-i207$ MeV
(also see section~\ref{Swavestudy}),
the $\rho$ pole at $756-i66.5$ MeV, corresponding
to a central resonance peak at 772 MeV and a width of 151 MeV, and the
$K^{\ast}$ pole at $892-i24.8$ MeV, resulting in
a peak at 895 MeV with a width of 52 MeV.

\section{Conclusions}
\label{conclusie}

The spectrum entering the RSE amplitudes as input is that of
a confined  (``quenched'') quark-antiquark pair.
In the present work, this spectrum is parametrised by the non-strange and
strange effective quark masses $m_{n}$ and $m_{s}$, and the strong
oscillator frequency $\omega$, all taken from Ref.~\cite{PRD27p1527}.
The mesonic resonances and their three-meson vertices are fully generated
by the RSE, and not input as in related yet perturbative methods.

The predictive power of the RSE as an analytic method to unquench the
quark model has been demonstrated before, by interrelating an enormous variety
of non-exotic mesonic systems, such as the light scalar mesons
$f_{0}$(600), $f_{0}$(980), $K_{0}^{\ast}$(800), $a_{0}$(980) and the
corresponding $S$-wave $\pi\pi$, $K\pi$, $\eta\pi$ scattering
observables \cite{ZPC30p615,PLB641p265}, the scalars between 1.3 and
1.5 GeV \cite{ZPC30p615}, vector and pseudoscalar mesons \cite{PRD27p1527},
charmonium and bottomonium \cite{PRD21p772}, the $D_{s0}^{\ast}$(2317)
\cite{PRL91p012003}, and the $D_{sJ}^{\ast}$(2860) \cite{PRL97p202001}.
In the present work, we have shown that also for $\pi\pi$ and $K\pi$
production, i.e., in three-body decay processes,
the RSE is a very powerful method to describe and understand the experimental
data.

\section*{Acknowledgements}

We wish to thank E.~Oset for outlining to us a simple method
on how to relate our elastic amplitudes to production processes.
We are furthermore indebted to D.~V.~Bugg for
carefully reading our manuscript,
suggesting improvements on the presentation, and advice
regarding the random generation of production data
as well as the handling of Dalitz plots.

This work was supported in part by the {\it Funda\c{c}\~{a}o para a
Ci\^{e}ncia e a Tecnologia} \/of the {\it Minist\'{e}rio da Ci\^{e}ncia,
Tecnologia e Ensino Superior} \/of Portugal, under contract
PDCT/\-FP/\-63907/\-2005.

\newcommand{\pubprt}[4]{#1 {\bf #2}, #3 (#4)}
\newcommand{\ertbid}[4]{[Erratum-ibid.~#1 {\bf #2}, #3 (#4)]}
\def\AIPCP{AIP Conf.\ Proc.}
\def\AP{Annals Phys.}
\def\EPJC{Eur.\ Phys.\ J.\ C}
\def\IJTPGTNO{Int.\ J.\ Theor.\ Phys.\ Group Theor.\ Nonlin.\ Opt.}
\def\JPG{J.\ Phys.\ G}
\def\MPLA{Mod.\ Phys.\ Lett.\ A}
\def\NPB{Nucl.\ Phys.\ B}
\def\PLB{Phys.\ Lett.\ B}
\def\PM{Phil.\ Mag.}
\def\PPNP{Prog.\ Part.\ Nucl.\ Phys.}
\def\PRC{Phys.\ Rev.\ C}
\def\PRD{Phys.\ Rev.\ D}
\def\PRL{Phys.\ Rev.\ Lett.}
\def\ZPA{Z.\ Phys.\ A}
\def\ZPC{Z.\ Phys.\ C}

\end{document}